\documentclass[twocolumn,superscriptaddress,showpacs,aps,prb]{revtex4}
\usepackage{amsmath}
\usepackage{amsfonts}
\usepackage{amssymb}
\usepackage{graphicx}
\usepackage{subfigure}
\usepackage{txfonts}
\usepackage{bm}
\usepackage[colorlinks, linkcolor=blue, anchorcolor=blue, citecolor=blue,
urlcolor=blue, dvipdfmx]{hyperref}

\begin{document}

\title{Exciton vortices in two-dimensional hybrid perovskite monolayers}

\begin{abstract}
We study theoretically the exciton Bose-Einstein condensation and exciton
vortices in a two-dimensional (2D) perovskite (PEA)${_2}$PbI${_4}$
monolayer. Combining the first-principles calculations and the Keldysh
model, the exciton binding energy of (PEA)${_2}$PbI${_4}$ in a monolayer can
approach hundreds meV, which make it possible to observe the excitonic
effect at room temperature. Due to the large exciton binding energy, and hence the high density of excitons, we find that the critical temperature of the
exciton condensation could approach the liquid nitrogen regime. In
presence of perpendicular electric fields, the dipole-dipole interaction
between excitons is found to drive the condensed excitons into patterned vortices,
as the evolution time of vortex patterns is comparable to the exciton
lifetime.
\end{abstract}

\author{Yingda Chen}
\affiliation{SKLSM, Institute of Semiconductors, Chinese Academy of Sciences, P.O. Box 912,
Beijing 100083, China}
\affiliation{CAS Center for Excellence in Topological Quantum Computation, University of
Chinese Academy of Sciences, Beijing 100190, China}
\author{Dong Zhang}
\email{zhangdong@semi.ac.cn}
\affiliation{SKLSM, Institute of Semiconductors, Chinese Academy of Sciences, P.O. Box 912,
Beijing 100083, China}
\affiliation{CAS Center for Excellence in Topological Quantum Computation, University of
Chinese Academy of Sciences, Beijing 100190, China}
\author{Kai Chang}
\email{kchang@semi.ac.cn}
\affiliation{SKLSM, Institute of Semiconductors, Chinese Academy of Sciences, P.O. Box 912,
Beijing 100083, China}
\affiliation{CAS Center for Excellence in Topological Quantum Computation, University of
Chinese Academy of Sciences, Beijing 100190, China}
\affiliation{Beijing Academy of Quantum Information Sciences, Beijing 100193, China}
\maketitle

Excitons, the composed bosons formed by bound electron-hole pairs through
Coulomb interactions, may collapse into Bose-Einstein condensation (BEC)
states at low temperatures\cite{snoke2,butov,intro5}. Bosons at the BEC
regime not only show exotic superfluity\cite{helium}, but also possess
patterns of vortices\cite{yarmchuk,keeling}. Such phenomena have been
studied both theoretically and experimentally recently in solids\cite%
{hqv,roy}, such as two-dimensional transitional metal dichalcogenides\cite%
{cyd}. To realize the exciton BEC, one need to find a system with long
exciton lifetime, huge binding energy and small exciton mass. Long exciton
radiative lifetime allows the excitons to build up a quasi-equilibrium
before recombination, while the huge binding energy leads to small Bohr
radius of excitons with high average exciton density. The 2D perovskite
monolayers could offer us a possible platform, promising considerably high
critical temperature of the exciton BEC.

In the past decades, hybrid organic-inorganic lead halide perovskites have
achieved remarkable records in the field of solar cells\cite%
{kojima,you,zhou,wanyingzhao}, and shown immense potentials as low-cost
alternatives to traditional semiconductors in commercial photovoltaic
industry\cite{jena,nayak}. Comparing to the three-dimensional (3D)
perovskites, 2D layered hybrid perovskites possess superior environment
stability in device performances\cite{tennyson,wenrongxie}, and provide
versatile blocks in dimensionality engineering\cite{grancini,quan} of
multi-dimensional perovskites due to the structural diversity\cite{li,lee}.
Surprisingly, the 2D hybrid perovskites display huge exciton binding
energies about hundreds meV\cite{exp,hong,yaffe} and long exciton lifetimes
about 2.5 ns\cite{fanghonghua}, even in presence of high defects and
disorders, due to the quantum confinement effects. The two distinguished
features make 2D hybrid organic-inorganic lead halide perovskites monolayers
as ideal platforms to realize exciton BEC.

\begin{figure}[tb]
\centering
\includegraphics[width=1.0\linewidth]{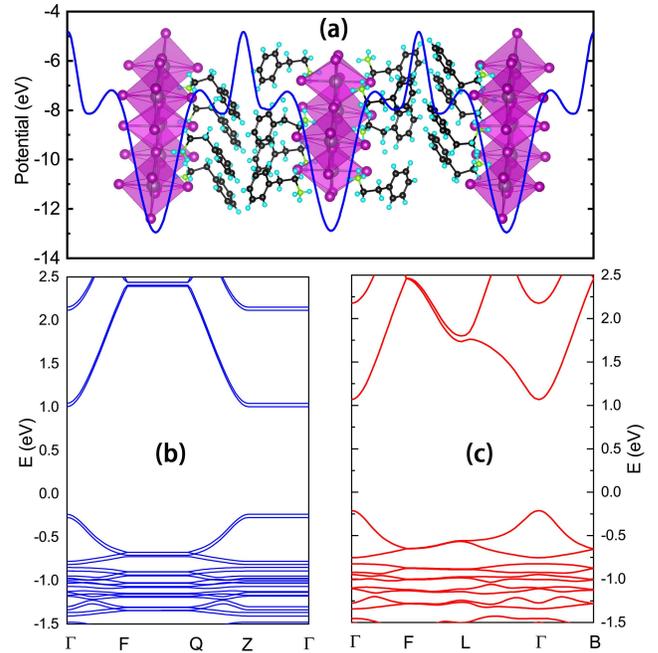}
\caption{(a) Schematic of layered hybrid perovskite (PEA)$_{2}$PbI$_{4}$,
for simplicity, the inequivalent two layers are labelled as Layer A and
Layer B. The different atoms are indicated by different colour coding, and
the effective inner potential is shown by blue solid lines. The lead,
iodide, carbon, hydrogen and nitrogen atoms are displayed by silver, purple,
black, cyan and green spheres, respectively. (b) Band structures of bulk
(PEA)$_{2}$PbI$_{4}$. (c) Band structures of (PEA)$_{2}$PbI$_{4}$ monolayer.}
\label{Fig1}
\end{figure}

In this work, we focus on the typical 2D hybrid perovskite (PEA)$_{2}$PbI$%
_{4}$\cite{hong,smith2,du,do}. The (PEA)$_{2}$PbI$_{4}$ possesses a stable
layered structure, which comprises alternatively stacked layers of [PbI$_{6}$%
]$^{4-}$ octahedra and long-chain organic molecules C$_{6}$H$_{5}$C$_{2}$H$%
_{4} $NH$_{3}^{+}$ (PEA$+$) as shown in Fig.~\ref{Fig1}\textcolor{blue}{(a)}%
. To determine both the crystalline structures and electronic structures,
the first-principles calculations are performed by using the Vienna ab
initio simulation package (VASP) within the generalized gradient
approximation (GGA) in Perdew-Burke-Ernzerhof (PBE) type and the projector
augmented-wave (PAW) pseudopotential. The kinetic energy cutoff is set to
500 eV for wave-function expansion, and the Monkhorst-Pack type k-point grid
is sampled by sums over 3$\times$3$\times$3. For the convergence of the
electronic self-consistent calculations, the total energy difference
criterion is set to 10$^{-8}$ eV. The crystal structure is fully relaxed
until the residual forces on atoms are less than 0.01 eV/\AA . The
spin-orbital coupling effect is taken into consideration, and the van der
Waals correction is also included by DFT-D2 method.

From Fig.~\ref{Fig1}\textcolor{blue}{(a)}, one can see that the inorganic
layers PbI$_{4}$ are sandwiched between two organic layers, with the
effective potential barriers as high as 8.1 eV, as illustrated by the blue
solid curves. Such high potential barriers make (PEA)$_{2}$PbI$_{4}$ behaves
like stacking quantum wells with hard-wall confining potentials. Due to the
weak interlayer Van der Waals coupling, we find the electronic structures
are similar between the bulk material and its monolayer. The band structures
are shown in Fig 1(c) all over high symmetric reciprocal points indicated in
the Brillouin zone. The parabolic conduction and valence bands are isolated
with bulk bands, and possess a direct band gap estimated as 1.24 eV. The
direct band gap feature ensures good performance of (PEA)$_{2}$PbI$_{4}$ in
optoelectronic devices. Notice that the band structures of the (PEA)$_{2}$PbI%
$_{4}$ monolayer (See Fig.~\ref{Fig1}\textcolor{blue}{(c)}) possess a direct
band gap about 1.278 eV at the $\Gamma$ point. Unlike other 2D materials
whose band gap vary significantly as the thickness decrease to monolayer,
e.g., the black phosphorous. The dimensionality reduction from bulk to the
monolayer limit does not increase the band gap evidently, since the
individual monolayer is naturally well confined by the internal potential
barriers, and the interlayer Van der Waals coupling is quite weak.

\begin{figure}[tb]
\centering
\includegraphics[width=1.0\linewidth]{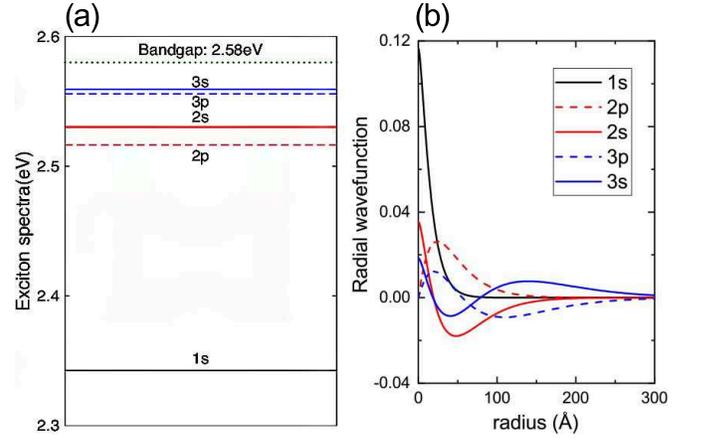}
\caption{(a) Exciton spectrum in perovskite monolayer estimated by the
Keldysh model. The solid lines show the s states, while the dash lines show
the p states. The dot green line gives the free carrier limit, which is
determined by the band gap of the 2D perovskite. (b) The schematic of radial
distribution of exciton wavefunctions.}
\label{Fig2}
\end{figure}

Based on the band structures obtained from the first-principles calculations
above, we study the excitons in (PEA)$_{2}$PbI$_{4}$ monolayer. The internal
motion of exciton is governed by%
\begin{equation}
\left[ -\frac{\hbar ^{2}}{2\mu }\nabla _{\boldsymbol{\rho }}^{2}+V_{2D}(%
\boldsymbol{\rho })\right] \psi =\mathcal{E} \psi .  \label{int}
\end{equation}%
where the reduced mass $\mu =m_{e}m_{h}/(m_{e}+m_{h})$, the electron mass $%
m_{e}=0.208m_{0}$, and the hole mass $m_{h}=0.372m_{0}$. The electron mass
and hole mass are adopted from the first-principles calculated band
dispersions along $\Gamma $-L path in k-space, respectively. Considering the
ultrathin thickness of the (PEA)$_{2}$PbI$_{4}$ monolayer, the Keldysh
potential\cite{keldysh} can be used to describe the Coulomb interaction
between the electron and hole.
\begin{equation}
V_{2D}\left( \rho \right) =-\frac{e^{2}}{4\pi \epsilon _{0}\epsilon _{2}\rho
_{0}}\frac{\pi }{2}\left[ H_{0}\left( \frac{\rho }{\rho _{0}}\right)
-Y_{0}\left( \frac{\rho }{\rho _{0}}\right) \right] ,  \label{scr}
\end{equation}%
where $\boldsymbol{\rho }=(\rho ,\varphi )$ is the relative displacement
between the electron and hole, the screening length of the PbI$_{4}$ layer $%
\rho _{0}=r_{0}/\epsilon _{2}$, and $r_{0}=\epsilon _{1}L_{w}(1+\epsilon
_{2}/\epsilon _{1}^{2})/2$. The relevant parameters in the Keldysh potential
are defined as follows. For the sandwiched inorganic PbI$_{4}$ layer, the
width $L_{w}=6.36${\AA }\cite{hong}, and the relative permittivity $\epsilon
_{1}=6.10$\cite{hong,exp}. For the organic barriers, the width $L_{b}=9.82${%
\AA }, and the relative permittivity $\epsilon _{2}=3.32$\cite{hong}. The
exciton binding energy $\mathcal{E} _{b}$ can be obtained by applying the
variational method to Eq.\eqref{int}. With the parameters given above, the
exciton binding energy is about $\mathcal{E} _{b}=$238.5 meV, which agrees
well with the experimental result 220-250meV\cite{hong,cheng}.
Correspondingly, the exciton spectrum of the (PEA)$_2$PbI$_4$ monolayer (the
red solid lines) is shown in Fig.~\ref{Fig2}\textcolor{blue}{(a)}. The
exciton spectrum is calculated from Eq.\eqref{int}. The bandgap $E_{g}=2.58$%
eV as reported in Ref.\cite{hong}, and the PL maximum $E_{0}=E_{g}-\mathcal{E%
} _{b}\simeq $2.34eV, which is close to the experimental value 2.4eV\cite%
{hong} and 2.37eV\cite{du}.

When a perpendicular electric field is applied, the effective electron-hole
interaction becomes dependent on the $z$-directional distribution of the
electron ($z_{e}$) and hole ($z_{h}$). The Keldysh potential takes the form
as
\begin{equation}
V\left( \rho ,z_{e},z_{h}\right) =-\frac{e^{2}}{4\pi \epsilon _{0}\rho }%
\int_{0}^{\infty }\frac{J_{0}\left( t\right) }{\epsilon _{mac}^{2D}\left(
t/\rho ,z_{e},z_{h}\right) }\mathrm{d}t,  \label{phi2d}
\end{equation}%
where the dielectric function is expressed as
\begin{equation}
\epsilon _{mac}^{2D}\simeq e^{q|z_{e}-z_{h}|}\left[ \epsilon _{2}+L_{w}q%
\frac{\epsilon _{1}}{2}\left( 1+\frac{\epsilon _{2}^{2}}{\epsilon _{1}^{2}}%
\right) \right] .  \label{emacu1}
\end{equation}%
in the thin film limit. By expanding the equation of exciton motion Eq.%
\eqref{int} into three dimensional(3D) form, the exciton motions under the
electric field are obtained,
\begin{equation}
\left[ -\frac{\hbar ^{2}}{2\mu }\nabla _{\boldsymbol{\rho }}^{2}-\frac{\hbar
^{2}}{2M}\nabla _{\mathbf{R}}^{2}+H_{ez}(z_{e})+H_{hz}(z_{h})+V\left(
\rho ,z_{e},z_{h}\right) \right] \psi =\mathcal{E}\psi .  \label{htot}
\end{equation}%
Here $\mathbf{R}$ is the displacement of the center of mass (c.m.) of
the exciton, $M=m_{e}+m_{h}$ is the c.m. mass of the exciton. $%
H_{ez}(H_{hz}) $ denotes the single-particle Hamiltonian of the
electron(hole) in the inorganic layer under the external electric field $F$. Variational exciton wavefunction\cite{miller2}, associated with the $l_{h}$%
th electron and $l_{e}$th hole subbands,
\begin{equation}
\psi _{nm}^{l_{e}l_{h}}(\rho ,z_{e},z_{h})=N^{l_{e}l_{h}}\zeta
_{l_{e}}\left( z_{e}\right) \zeta _{l_{h}}\left( z_{h}\right) e^{-\sqrt{%
\left( \frac{z_{e}-z_{h}}{z_{0}}\right) ^{2}+\left( \frac{\rho }{a_{0}}%
\right) ^{2}}}  \label{phitot}
\end{equation}%
is adopted for the 1$s$-type state in Eq.\eqref{htot}, with variational
parameters $a_{0}$ and $z_{0}$. The $z$-directional confinements of the
electron and hole are included in $\zeta _{l_{e}}$ and $\zeta _{l_{h}}$ (See
APPENDIX~\ref{solution}).

\begin{figure}[tb]
\centering
\includegraphics[width=1.0\linewidth]{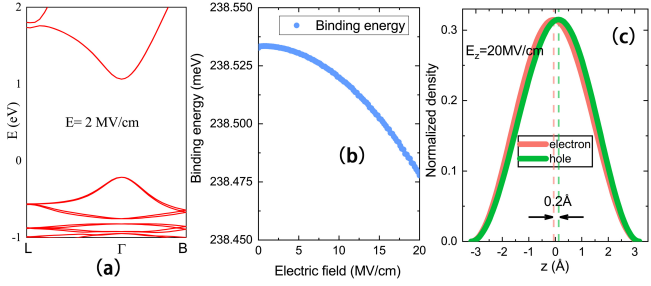}
\caption{(a) Band structures of (PEA)$_{2}$PbI$_{4}$ monolayers under
external electric field of 2 MV/cm. (b) The exciton binding energies and
strengths of perpendicular electric fields. (c) The electron-hole spatial
separation as a function of perpendicular electric field. The effective
separation is denoted by distance between centres of electrons (red lines)
and holes (green lines).}
\label{Fig3}
\end{figure}

For the 1$s$-type exciton state, the effects of perpendicular electric
fields are limited. As shown in Fig~\ref{Fig3}\textcolor{blue}{(a)}, The
band gap of the perovskite monolayer remains unchanged under an electric
field 2 MV/cm. Accordingly, the binding energy of the exciton varies
slightly. From Fig~\ref{Fig3}\textcolor{blue}{(b)}, one can find the binding
energy drops merely 0.06 meV as the strength of the perpendicular electric
field increases up to a very strong electric field 20 MV/cm. The reason lies
in the fact that, it is difficult to separate electrons and holes
vertically, as shown in Fig~\ref{Fig3}\textcolor{blue}{(c)}, in presence of
perpendicular electric field as strong as 20 MV/cm, the effective spatial
separation between centres of holes and electrons $d$ along the z-axis is less
than 1 $\mathring{A}$. This feature arises from the very strong confining
potential (about 8eV) and weak inter-layer coupling.

Although perpendicular electric field changes extremely slightly the binding
energies and Bohr radius of the perovskite monolayer, but it is efficient to
align dipoles, change the exciton-exciton interactions utterly, and play a
crucial role in the exciton BEC under the critical temperature.

The critical temperature for the BEC transition in the flakes of (PEA)$_{2}$%
PbI$_{4}$ monolayer is estimated by
\begin{equation}
T_{c}=\frac{\pi \hbar ^{2}n}{Mk_{B}}\frac{1}{\ln (nS/2)}
\end{equation}%
where $S$ is the area of the flake, and $n$ is the exciton density. For a
square flake, with $S=$(20nm)$^{2}$ and $n=5\times $10$^{12}$cm$^{-2}$, we
obtain $T_{c}\simeq 106$K. Therefore, the exciton condensation can be
achieved at the temperature under liquid nitrogen regime in (PEA)$_{2}$PbI$%
_{4}$ monolayer.

Usually, the Gross-Pitaevskii (GP) equation is widely used to describe the
condensate states. Considering laser pumping and exciton recombination
process, the non-equilibrium exciton condensates in the perovskite monolayer
with the lateral boundaries, can be described by complex Gross-Pitaevskii
(cGP) equation within mean-field approach. Under a weak perpendicular
electric field, the exciton-exciton interaction is dominated by repulsive
dipole-dipole interaction (DDI), which can be expressed as $V_{dd}$ in
reciprocal space,
\begin{equation}
V_{dd}\left( \mathbf{Q}\right) =\frac{e^{2}d}{\epsilon _{0}\epsilon _{1}}%
\frac{2\left[ (1+\frac{\epsilon _{2}}{\epsilon _{1}})e^{x_{Q}}+(1-\frac{%
\epsilon _{2}}{\epsilon _{1}})e^{-x_{Q}}\right] -4}{x_{Q}[(1+\frac{\epsilon
_{2}}{\epsilon _{1}})^{2}e^{x_{Q}}-(1-\frac{\epsilon _{2}}{\epsilon _{1}}%
)^{2}e^{-x_{Q}}]},
\end{equation}%
with $x_{Q}=Qd$. Here $d$ is the electron-hole separation introduced
by the electric field. Thus the cGP equation is expressed as
\begin{equation}
i\hbar \partial _{t}\psi =\left[ -\frac{\hbar ^{2}}{2M}\nabla
^{2}_{\mathbf{R}}+V_{c}+H_{dd}+i\hbar \left( \hat{R}-\Gamma \left\vert \psi
\right\vert ^{2}\right) \right] \psi ,  \label{cgpe}
\end{equation}%
where $\hat{R}$ is the pumping rate, $\Gamma =1/2\tau _{ex}$ the
recombination rate, and $\tau _{ex}$ the exciton lifetime. $V_{c}$ denotes
as the trap potential for excitons. It is worth noting that,
convolution of $H_{dd}=V_{dd}\ast \left\vert \psi \right\vert ^{2}$
stands for the DDI, which display a nonlinear behavior.

Considering the laser pumping is generally radial symmetric, while the
perovskite flakes possess irregular shapes in practice, the disorders induced by
the lateral boundary provide the scatterings to the optically generated excitons,
and change the directions of their momentum. As a consequence,
non-zero angular velocities appear at the edges of the flakes. In presence of
the nonlinear DDI term of the flake boundaries, the exciton condensate state is sensitive to the local angular velocities raised by the lateral disorders, which behave as local potentials surrounding the condensate cloud. Therefore, the exciton vortices are expected to emerge in the perovskite monolayer flakes with constant laser pumping.

In order to demonstrate the vortex states, the complex GP equation\eqref{cgpe} is solved by time-splitting spectral methods in combination with discrete sine transforms\cite%
{yycai,yycai2,yycai3,sierra}. The pumping and decaying process in each time
step is shown in APPENDIX~\ref{pump}. The parameters for exciton vortex
simulations in a square flake of (PEA)$_{2}$PbI$_{4}$ monolayer with length $%
L_{0}$=400nm, are listed as follows. The flake is sampled by a 512 $\times$
512 mesh in real space, and the time step is set to be 5$\times 10^{-6}$%
ns, to reach high accuracy. The pump power is set to be $\hat{R}\simeq
2\times 10^{-3}$meV, and the exciton lifetime is $\tau _{ex}$=2ns. A shallow harmonic
potential is also introduced to mimic interface potential fluctuations in the quantum well structures, i.e., we set $V_{c}=V_{0}R^{2}/2L_{0}^{2}$ for $R<L_{0}$ and $V_{c}=\infty$ for $R\geq L_{0}$, with $V_{0}\sim $10meV and $R=|\mathbf{R}|$.

\begin{figure}[tb]
\centering
\includegraphics[width=1.0\linewidth]{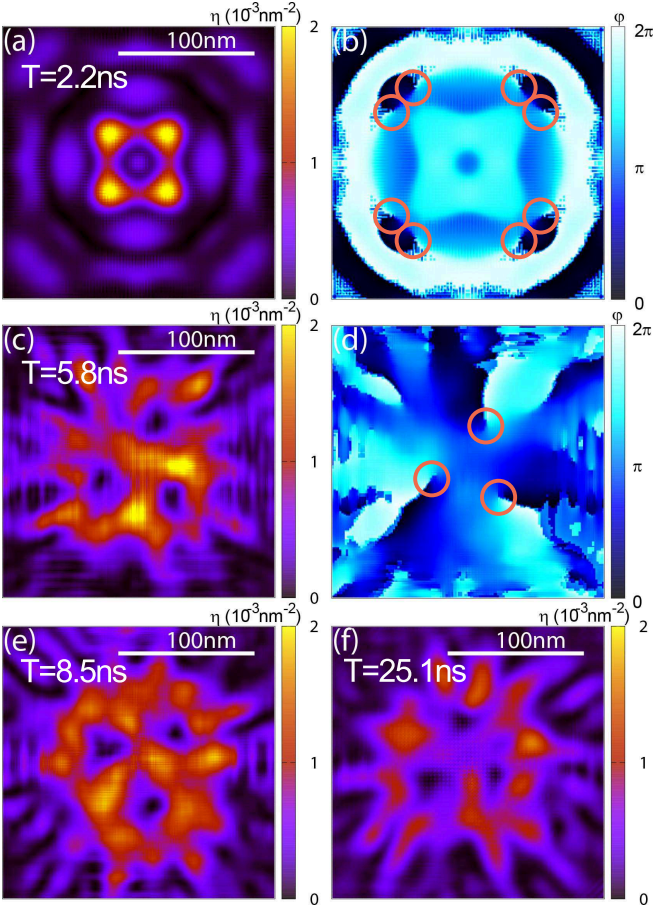}
\caption{Dynamic evolution of exciton vortices patterns in perovskite
monolayer flake. (a) Contour map of the density of the exciton condensates
wavefunction, at the first occurrence of occasional exciton vortices
pattern. (b) The phase distributions of the exciton condensates
wavefunction. (c) and (d), Density and phase distributions of exciton
condensates wavefunction of the stable vortices pattern. (e) and (f),
Density of exciton condensates wavefunction of the stable vortices pattern
at T = 8.5 ns and T = 25.1 ns, respectively. The density $\protect\eta=|\psi|^{2}$ is
in the unit of $10^{-3}$nm$^{-2}$. The orange circles show the locations
of the vortices.}
\label{Fig4}
\end{figure}

The simulated exciton vortices are shown in Fig.~\ref{Fig4}. The vortex is
characterized by a rotation of the phase of the condensates wavefunction
around the singular point by an integer multiple of 2$\pi$. As shown in
Figure~\ref{Fig4}\textcolor{blue}{(a)}, the first occasional vortices emerge
during the dynamic evolution of the exciton condensates at T = 2.2 ns. The
eight vortices locate at the dark spots in the contour plot of the density
of the exciton condensates wavefunction. The corresponding phase
distributions in Fig.~\ref{Fig4}\textcolor{blue}{(b)}, indicate 2$\pi$ phase
shift around the singular points of the wavefunction, which indicate the
existence of exciton vortices. As time goes by, the vortex patterns tend to
reach dynamic equilibrium since T = 5.8 ns. From Fig.~\ref{Fig4}%
\textcolor{blue}{(c)} and Fig.~\ref{Fig4}\textcolor{blue}{(d)}, one can find
there are three vortices in the central area of the perovskite monolayer
flake. The three vortices are stable and rotating persistently, as
illustrated at T = 8.5 ns, and even T = 25.1 ns, respectively, as shown in
Fig.~\ref{Fig4}\textcolor{blue}{(e)} and Fig.~\ref{Fig4}\textcolor{blue}{(f)}%
. Since the evolution time of vortices patterns is comparable to the exciton
lifetime, it is promising to observe exciton vortex patterns in (PEA)$_2$PbI$%
_4$ monolayers experimentally.

In summary, we study the exciton BEC and its vortices in (PEA)$_2$PbI$_4$
monolayer, and calculated exciton binding energy 238.5 meV is in good
agreement with experimental results. We find the perpendicular electric
fields change slightly the binding energy and Bohr radius in (PEA)$_2$PbI$_4$
monolayer, but are efficient to align the electron-hole dipoles. With laser
pumping, the repulsive dipole-dipole interaction created by the
perpendicular electric field can drive the laterally confined excitons into
various vortex patterns. The evolution time of those vortices is comparable
to the exciton lifetime, and reach a stable pattern with certain number of
vortices rotating at the center. Since the large exciton binding energy
ensures the critical temperature of the exciton BEC within the liquid
nitrogen regime, it is possible to realize stable exciton vortices in
two-dimensional hybrid perovskite monolayers.

This work was supported by National Key R$\&$D Programmes of China, Grant
No. 2017YFA0303400, 2016YFE0110000. National Natural Science Foundation of
China, Grant No. 11574303, 11504366. Youth Innovation Promotion Association
of Chinese Academy of Sciences, Grant No. 2018148, and the Strategic
Priority Research Program of Chinese Academy of Sciences, Grant No.
XDB28000000.

\appendix

\section{Solution of the quantum confined exciton under electric field}

\label{solution} The single particle $H_{ez}(H_{hz})$ denotes the
Hamiltonian of the electron(hole) in the inorganic layer under the electric
field $F$, i.e.,%
\begin{equation}
\left\{
\begin{array}{c}
H_{ez}(z_{e})=-\frac{\hbar ^{2}}{2m_{e}}\frac{\partial ^{2}}{\partial
z_{e}^{2}}+V_{\text{conf}}\left( z_{e}\right) +eFz_{e} \\
H_{hz}(z_{h})=-\frac{\hbar ^{2}}{2m_{h}}\frac{\partial ^{2}}{\partial
z_{h}^{2}}+V_{\text{conf}}\left( z_{h}\right) -eFz_{h}%
\end{array}%
\right. ,  \label{hz}
\end{equation}%
where $V_{\text{conf}}$ is the infinite-potential-barrier for the electron
and the hole, since they are confined in the single inorganic layer.
Eigenstates of Eq.\eqref{hz} are represented by the Airy functions \textrm{Ai%
} and \textrm{Bi},
\begin{equation}
\zeta _{l_{e}(l_{h})}\left( z\right) =a_{l_{e}(l_{h})}\mathrm{Ai}\left(
Z_{l_{e}(l_{h})}(z)\right) +b_{l_{e}(l_{h})}\mathrm{Bi}\left(
Z_{l_{e}(l_{h})}(z)\right) ,  \label{phiz}
\end{equation}%
where $l_{e}(l_{h})=1,2,\ldots $ is the subband index of the electron(hole),
and
\begin{equation}
\left\{
\begin{array}{l}
Z_{l_{e}}=-[2m_{e}/\left( e\hbar F\right) ^{2}]^{\frac{1}{3}%
}(E_{l_{e}}^{(e)}-eFz) \\
Z_{l_{h}}=-[2m_{h}/\left( e\hbar F\right) ^{2}]^{\frac{1}{3}%
}(E_{l_{h}}^{(h)}+eFz)%
\end{array}%
\right. .
\end{equation}%
Here $E_{l_{e}}^{(e)}(E_{l_{h}}^{(h)})$ is the $l_{e}$-th electron ($l_{h}$%
-th hole) subband energy. The parameters are normalized as%
\begin{equation*}
\left\{
\begin{array}{l}
a_{l}=[1+\mathrm{Ai}^{2}\left( Z_{l\pm }\right) /\mathrm{Bi}^{2}\left(
Z_{l\pm }\right) ]^{-1/2} \\
b_{l}=-a_{l}\mathrm{Ai}\left( Z_{l\pm }\right) /\mathrm{Bi}\left( Z_{l\pm
}\right)%
\end{array}%
\right.
\end{equation*}%
, with $Z_{l\pm }=Z_{l}(z=\pm L_{w}/2)$. The eigenenergies $E_{l}$\ are
determined by the $(l-1)$-zeros of%
\begin{equation}
S\left( E_{l}\right) =\mathrm{Ai}\left( Z_{l+}\right) \mathrm{Bi}\left(
Z_{l-}\right) -\mathrm{Bi}\left( Z_{l+}\right) \mathrm{Ai}\left(
Z_{l-}\right) .  \label{se}
\end{equation}

\section{Treatment of the pumping and decaying term}

\label{pump} Here we deal with the time step containing the pumping and
decaying term $i\partial _{t}\psi =i\left( \hat{R}-\Gamma \left\vert \psi
\right\vert ^{2}\right) \psi $, which can be written as%
\begin{equation}
\partial _{t}\psi =\left( \hat{R}-\Gamma \left\vert \psi \right\vert
^{2}\right) \psi .  \label{phi}
\end{equation}%
Since the Eq.\eqref{phi} is real, we have%
\begin{equation}
\partial _{t}\rho =2\left( \hat{R}-\Gamma \rho \right) \rho .  \label{rho}
\end{equation}%
Next we solve $\rho \left( t\right) $ from Eq.\eqref{rho}, and obtain%
\begin{equation}
\rho \left( t\right) =\frac{\hat{R}}{\Gamma }\frac{1}{1+Ce^{-2\hat{R}t}},
\label{rhot}
\end{equation}%
with%
\begin{equation*}
C=\frac{\hat{R}}{\Gamma }\frac{1}{\rho _{0}}-1.
\end{equation*}%
Recalling Eq.\eqref{phi},
\begin{equation}
\psi _{1}^{\prime }=e^{\int_{0}^{\Delta t}\left( \hat{R}-\Gamma \rho \left(
s\right) \right) ds}\psi _{1},\rho \left( 0\right) =\psi _{1}^{2}.
\label{phie}
\end{equation}%
Putting Eq.\eqref{rhot} into the integral in Eq.\eqref{phie}, we have
\begin{equation}
\int_{0}^{\Delta t}\left( \hat{R}-\Gamma \rho \left( s\right) \right) ds=%
\hat{R}\Delta t+\frac{1}{2}\ln \left( \frac{\hat{R}}{\hat{R}+\Gamma \rho
_{0}\left( e^{2\hat{R}\Delta t}-1\right) }\right)
\end{equation}%
The second-order time splitting can be written as%
\begin{equation}
\psi \left( t+\Delta t\right) =e^{-i\frac{T}{2}\Delta t}e^{-i\frac{V}{2}%
\Delta t}e^{\int_{0}^{\Delta t}\left( \hat{R}-\Gamma \rho \left( s\right)
\right) ds}e^{-i\frac{V}{2}\Delta t}e^{-i\frac{T}{2}\Delta t}\psi \left(
t\right).
\end{equation}%

\bibliographystyle{apsrev4-2}
\bibliography{ref}

\end{document}